# Second-Generation Heterogeneous Panel Data Model with Individual and Common Shocks

Hasraddin Guliyev[1]

[1] Azerbaijan State University of Economics, School of Master's Studies, Baku, AZ1007, Azerbaijan.
E-mail: hasradding@unec.edu.az

**Abstract**

We study estimation of the mean slope in heterogeneous panels that combine cross-sectional dependence from unobserved common factors with unit-specific structural breaks occurring at different dates. We organise the available second-generation Mean Group estimators into a regime map indexed by the cross-section size, the strength of the cross-sectional dependence, and the nature of the structural change, and we examine two estimators for the small-to-moderate-dependence panels common in applied macroeconomics and energy economics. The Fourier SUR Mean Group (F-SURMG) estimator augments a seemingly unrelated regression system with unit-specific Fourier terms. The proposed Fourier Common Correlated Effects Mean Group (F-CCEMG) estimator augments the CCE regression with deterministic Fourier terms, filtering the common factor while absorbing the heterogeneously-timed breaks. In a Monte Carlo study with R = 500 replications across weak, moderate, and strong dependence, F-CCEMG attains the lowest root mean squared error in almost every configuration and near-nominal coverage once the cross-section is not minimal, while F-SURMG gives the best-calibrated inference in the small-N, weak-dependence corner; estimators that do not filter the factor lose coverage as dependence rises. An application to the renewable energy–growth nexus in the G7 over 1965–2019 finds no significant aggregate effect of renewable-energy consumption on growth.

## 1. Introduction

A central question in applied panel econometrics is whether the relationship between economic variables is common to all cross-sectional units or whether it differs across them. The conventional pooled, fixed-effects (FE), and random-effects (RE) estimators rest on slope homogeneity: the marginal effect of the regressors is taken to be identical across units, and only the intercept is allowed to vary, either as a fixed unit-specific term (FE) or as a random draw from a common distribution (RE). This restriction is convenient, since it allows the data to be pooled and the common slope to be estimated from the full cross-section–time variation, delivering precise estimates when the restriction holds. When the slopes genuinely differ across units, however, the assumption is no longer innocuous. Pesaran and Smith (1995) show that in dynamic heterogeneous panels, pooling and aggregation yield inconsistent and potentially misleading estimates of the average relationship, because the imposed common slope confounds the heterogeneous responses with the dynamics of the regressors. The choice between homogeneous and heterogeneous specifications is therefore not merely a matter of efficiency but of consistency: where heterogeneity is present, FE and RE answer a different—and generally incorrect—question about the average effect of interest.

The first generation of heterogeneous estimators dispenses with the pooling restriction. The Mean Group (MG) estimator of Pesaran and Smith (1995) runs a separate time-series regression for each unit and averages the unit-specific coefficients to recover the mean of the heterogeneous slope distribution. MG is consistent under slope heterogeneity and requires no assumption about the form of that heterogeneity, which makes it the natural benchmark when both the cross-sectional dimension N and the time dimension T are moderately large. Its principal limitation is that it treats the units as cross-sectionally independent: the unit-by-unit regressions are valid only if the errors are uncorrelated across units, so that averaging does not transmit a common disturbance into the group mean. In macroeconomic and energy-economics panels this independence is rarely tenable, because countries are exposed to common global shocks—oil-price movements, financial crises, technological waves, coordinated policy regimes—that induce strong cross-sectional dependence (CSD). When such common factors are present and correlated with the regressors, the unit-level regressions underlying MG are themselves contaminated, and the group average inherits the bias (Pesaran, 2006).

The second generation of estimators was developed to address this problem. Pesaran (2006) proposes the Common Correlated Effects (CCE) approach, which augments each unit regression with the cross-sectional averages of the dependent variable and of the regressors. Under a general multifactor error structure, these averages span the space of the unobserved common factors asymptotically as N grows, so the differential effect of the factors is filtered out of each unit equation without the factors ever being estimated directly. The Common Correlated Effects Mean Group (CCEMG) estimator then averages the filtered unit slopes in the spirit of MG, combining robustness to slope heterogeneity with robustness to cross-sectional dependence. Pesaran (2006) reports satisfactory small-sample properties even under substantial heterogeneity, and the approach has been extended to dynamic panels with weakly exogenous regressors (Chudik and Pesaran, 2015) and to non-stationary multifactor structures (Kapetanios, Pesaran, and Yamagata, 2011). CCE has become the workhorse for panels with cross-sectional dependence and is the appropriate point of departure for any modern treatment of heterogeneous panels.

The validity of the cross-sectional-average approximation depends on two features of the data: the strength of the common factors and the dimensions of the panel. The averages are informative proxies for the latent factors only when the factor loadings do not average to zero and when N is large enough for the averaging to recover the factor space. CCE is therefore most powerful when both N and T are large and the dependence is strong, since the common component then dominates and the averages absorb it efficiently. This logic extends to structural change. When the breaks that affect the panel are driven by the common factors themselves—entering the data through the same channel as the dependence—the cross-sectional averages track those breaks by construction, even when their timing and magnitude differ across units. Under strong CSD the CCE augmentation thus absorbs not only the smooth common variation but also the heterogeneously-timed shifts that the common factors transmit to individual units. Baltagi, Feng, and Kao (2016) confirm this, showing that the CCE estimator attains the same asymptotic distribution under common breaks as it would were the break dates known, and Ditzen, Karavias, and Westerlund (2024) develop a complete inferential toolbox for multiple structural breaks in interactive-effects panels. The message is clear: when N and T are large and the common factors are strong, CCEMG handles unit-specific breaks that originate in the common component, and little is gained by modelling each unit's break separately.

The difficulty is that many panels of greatest interest in applied macroeconomics and energy economics do not inhabit this large-N, strong-CSD regime. Country groupings such as the G7, the BRICS, and the Next-Eleven (N-11) are deliberately small in the cross-section—often N below ten—while the time span is long, so T substantially exceeds N. In exactly this configuration the cross-sectional-average approximation is least reliable: with so few units the averages are noisy proxies for the latent factors, and the asymptotic argument that justifies CCE has little force. Moreover, the structural changes in such panels are frequently idiosyncratic rather than common. The G7 economies, for example, reached energy-transition turning points at very different dates—reflecting national policy shifts, the timing of nuclear or hydropower commitments, and country-specific responses to events such as the Fukushima accident—so that each unit breaks on its own schedule rather than in response to a single shared factor. When the dependence is weak or moderate and the breaks are individual, the cross-sectional averages can neither proxy a now-weak factor structure nor capture breaks unrelated to that structure. The CCE filter is then largely empty, and a different strategy is required: one that models each unit's structural change directly while still accommodating slope heterogeneity.

The Seemingly Unrelated Regressions Mean Group (SURMG) family answers this need. Building on the Mean Group principle, Guliyev (2023) estimates the system of unit-specific equations jointly as a Seemingly Unrelated Regression, exploiting the contemporaneous correlation across unit errors to gain efficiency precisely when N is small and T is large—the regime in which CCE struggles. To incorporate structural change without the loss of degrees of freedom and the multicollinearity that dummy-variable schemes entail in small samples, the Fourier SURMG (F-SURMG) estimator augments each unit equation with low-frequency trigonometric terms. The Fourier approach, introduced to the structural-break literature by Becker, Enders, and Lee (2006) and Enders and Lee (2012), approximates breaks of unknown number, timing, and form with a small number of sine and cosine components, and a single frequency is typically sufficient to capture smooth as well as sharp transitions. Because the Fourier terms are unit-specific, F-SURMG captures structural changes that occur at different dates in different units—exactly the idiosyncratic-break pattern that defeats the common-factor logic of CCE. Guliyev (2025) demonstrates the value of this approach in the G7 setting, where weak-to-moderate dependence coexists with country-specific transition dates and where F-SURMG recovers heterogeneous renewable-energy elasticities that pooled and common-factor estimators obscure. The SB-SURMG variant, which uses estimated break dummies, provides the sharp-break counterpart when the timing of individual breaks can be pinned down.

Taken together, this literature suggests that the appropriate estimator is dictated by the joint configuration of the panel dimensions, the strength of the cross-sectional dependence, and the nature of the structural change—yet no single study maps these regimes systematically or brings the estimators tailored to small, strongly heterogeneous panels into a common framework. This paper fills that gap. We organise the heterogeneous-panel estimators into a coherent regime map and, through a Monte Carlo design that varies N, T, the intensity of the common factor, and the dispersion of unit-specific break dates, we characterise where each estimator dominates. The methodological contribution is twofold. First, we develop the Fourier Seemingly Unrelated Regressions Mean Group (F-SURMG) estimator within this unified framework as the estimator of choice for the small-N, large-T, weak-dependence regime: by estimating the unit equations jointly as a Seemingly Unrelated Regression and augmenting each with unit-specific Fourier terms, F-SURMG exploits the cross-equation correlation for efficiency while capturing breaks at different dates across units. Second, we propose the Fourier Common Correlated Effects Mean Group (F-CCEMG) estimator, which augments the CCE unit regressions with deterministic Fourier terms before mean-group averaging. F-CCEMG is designed for the intermediate regime—moderate dependence combined with individual, heterogeneously-timed breaks—in which the averages still carry some factor information but cannot, on their own, absorb unit-specific structural change; here it combines the factor-filtering of CCE with the break-approximating flexibility of the Fourier form. We propose F-CCEMG and study its finite-sample behaviour by simulation rather than establishing its limiting distribution: the asymptotic argument is heuristic, inherited from CCE, and we treat the expected consistency and variance reduction as conjectures that the Monte Carlo evidence is designed to test. Together, the two estimators close the methodological gap between plain CCEMG, built for the large-N, strong-dependence regime, and the unmodelled small-sample heterogeneous-break case. The resulting regime map is straightforward and practically useful: when N and T are large and the dependence is strong, CCEMG already accommodates unit-specific breaks transmitted through the common factors and should be preferred; when N is small relative to T and the dependence is weak while breaks are individual, F-SURMG is the most reliable; and in the intermediate case of moderate dependence with idiosyncratic breaks, F-CCEMG offers the best performance. These distinctions bear directly on the empirical analysis of small, strongly heterogeneous country groups—the G7, the BRICS, and the N-11—where the assumption of strong common factors is often unwarranted but individual structural change is the rule rather than the exception.

The remainder of the paper is organised as follows. Section 2 sets out the heterogeneous panel data model with individual and common shocks, reviews the FE, MG, CCEMG, and F-SURMG estimators within a common framework, and introduces the proposed F-CCEMG estimator together with its asymptotic motivation. Section 3 reports the Monte Carlo design and results, establishing the regime map across the panel dimensions, the strength of cross-sectional dependence, and the dispersion of break dates. Section 4 presents an empirical application to a small heterogeneous country group, and Section 5 concludes with the methodological and policy implications.

## 2. The heterogeneous panel model with individual and common shocks

Consider a panel of N cross-sectional units observed over T periods. We work with the heterogeneous panel model

$$y_{it} = \alpha_i + \beta'_i x_{it} + u_{it}, \quad u_{it} = \gamma'_i f_t + \varepsilon_{it}, \quad i = 1, \ldots, N; \; t = 1, \ldots, T, \quad (1)$$

where $y_{it}$ is the outcome, $x_{it}$ is a k×1 vector of observed regressors, and αi is a unit-specific intercept. The slope $\beta_i$ is heterogeneous across units, drawn from a common distribution with mean $\beta = E(\beta_i)$; β is the parameter of primary interest—an estimand, the mean of the unit-level effects, not a structural causal parameter. The composite error uit has an idiosyncratic part $\varepsilon_{it}$ (mean zero, finite variance) and a common part $\gamma'_i$ ft, where $f_t$ is an m×1 vector of unobserved factors and $\gamma_i$ the heterogeneous loadings. The factors generate cross-sectional dependence; when they are correlated with $x_{it}$, the regressors are endogenous with respect to $u_{it}$.

The econometrician observes ($y_{it}$, $x_{it}$); the factors $f_t$, the loadings $\gamma_i$, the unit slopes $\beta_i$, and the errors are latent (Pesaran, 2006; Chudik and Pesaran, 2015). The object of estimation is the scalar-or-vector mean β; the unit slopes $\beta_i$ are nuisance quantities estimated only as an intermediate step.

Three assumptions delimit the model and separate its identifying content from the conditions needed only for asymptotic arguments. The slopes are independent draws from a distribution with finite mean and variance, $E(\beta_i) = \beta$ and $Var(\beta_i) = \Omega\beta$, which defines β as the estimand and makes the dispersion of the unit slopes a genuine feature of the data rather than sampling noise. The factors are bounded and their heterogeneous loadings have a non-zero average, so the common component does not vanish in the cross-section; this is what allows the cross-sectional averages used below to proxy the factors. The idiosyncratic errors are mean-zero with finite variance and are independent of the regressors once the factor and any deterministic terms are accounted for, while the regressors are correlated with the factor—the source of the endogeneity that distinguishes second-generation estimators from first-generation ones. These conditions are standard in the CCE literature (Pesaran, 2006; Chudik and Pesaran, 2015); the unit-specific level break added in the simulation enters through the idiosyncratic component and is, by design, orthogonal to the factor.

### 2.1. The fixed-effects and Mean Group estimators

The homogeneous fixed-effects (FE) estimator imposes a common slope, $\beta_i = \beta$ for all i, and pools the within-transformed data, estimating β from the entire cross-section–time variation. When the slopes are in fact equal this is the efficient choice. When they differ, however, the restriction is not merely inefficient but inconsistent for the mean slope β. Pesaran and Smith (1995) show that the pooled estimator converges to a weighted combination of the unit slopes in which the weights depend on the second moments of the regressors, so that units with more variable regressors are over-represented and the probability limit differs from the simple average E(βi); in dynamic settings the same mechanism mixes the heterogeneous responses with the dynamics of the regressors and can reverse signs. The estimator then answers a different question from the one of interest. The Mean Group (MG) estimator of Pesaran and Smith (1995) avoids this by abandoning the pooling restriction altogether: it estimates a separate time-series regression for each unit,

$$\hat{\beta}_i = (X'_i X_i)^{-1} X'_i y_i, \quad (2)$$

with $X_i$ and $y_i$ the within-unit data, and averaging the unit estimates,

$$\hat{\beta}_{MG} = N^{-1} \Sigma_i \hat{\beta}_i. \quad (3)$$

MG is consistent for $\beta$ under heterogeneity, with the non-parametric Pesaran–Smith variance $\text{var}(\hat{\beta}_{MG}) = [N(N-1)]^{-1} \Sigma_i(\hat{\beta}_i - \hat{\beta}_{MG})(\hat{\beta}_i - \hat{\beta}_{MG})'$. Its validity rests on cross-sectional independence: when $f_t$ is present and correlated with $x_{it}$, each $\hat{\beta}_i$ is biased and the average inherits the bias. Independence rules out common shocks—untenable in country panels—so MG is the cautionary benchmark rather than a serious competitor under dependence.

What this independence assumption rules out, and why it fails here, is worth stating directly. MG requires that, after conditioning on the regressors, the unit errors be uncorrelated across units, so that the sampling errors of the individual slope estimates are independent and their average concentrates on $\beta$. A common factor violates this directly: it makes the errors of every unit move together, and—more damagingly—when the factor also drives the regressors, it correlates each regressor with its own equation's error, so that every unit estimate $\hat{\beta}i$ is biased in the same direction and the bias does not average away. The failure is thus not one of efficiency that a larger sample would cure, but of consistency. Because common global shocks are the norm rather than the exception in country panels, we treat MG as a benchmark that isolates the cost of ignoring dependence, not as a serious candidate for the applications we have in mind.

## 2.2. The Common Correlated Effects Mean Group estimator

The Common Correlated Effects approach of Pesaran (2006) controls for the unobserved factors without estimating them. The key observation is that the cross-sectional averages of the observed variables are driven by the same factors as the individual series: averaging (1) across units, the idiosyncratic errors wash out as N grows while the common component $\bar{\gamma}'$ ft survives, so the averages are, asymptotically in N, exact linear combinations of the latent factors. Including these averages in each unit equation therefore proxies for the factors directly. Concretely, each unit regression is augmented with the cross-sectional averages of the dependent variable and the regressors, $\bar{z}_t = (\bar{y}_t, \bar{x}'_t)'$, giving the unit regression

$$y_{it} = \alpha_i + \beta'_i x_{it} + \delta'_i \bar{z}_t + e_{it} \quad (4)$$

spans and removes the factor space. The CCE Mean Group (CCEMG) estimator averages the resulting slopes, $\hat{\beta}_{CCEMG} = N^{-1} \Sigma_i \hat{\beta}_{i,CCE}$, with the same non-parametric variance as MG, and is consistent for $\beta$ as N, $T \to \infty$. Two qualifications matter here. First, the spanning argument is asymptotic in N; when N is very small the averages are noisy proxies and the filter is imprecise. Second, the averages absorb only common variation: breaks that are idiosyncratic—at different dates across units and orthogonal to $f_t$—remain in $e_{it}$, inflating its variance and the dispersion of the unit slopes. These two qualifications are the openings the proposed estimator exploits.

The framework has been extended well beyond the static case. Chudik and Pesaran (2015) adapt the CCE approach to dynamic heterogeneous panels with weakly exogenous regressors by adding lags of the cross-sectional averages, and Kapetanios, Pesaran, and Yamagata (2011) show that the cross-sectional averages continue to span the factor space when the common factors are non-stationary, so that the estimator remains valid under unit-root factors. These developments have made CCE the standard tool for heterogeneous panels with cross-sectional dependence. The present paper stays within the static specification, since the renewable-energy–growth relationship of the application is modelled in levels, but the Fourier augmentation introduced below is compatible with these extensions in principle.

The behaviour of CCE under structural change depends on how the breaks enter the data. When the breaks are transmitted through the common factors—so that a shift in ft is felt by every unit according to its loading—the cross-sectional averages move with the breaks by construction and absorb them automatically, even when the resulting shifts occur at different dates and with different magnitudes across units. In this case the literature shows that little is lost by not modelling the breaks explicitly: Baltagi, Feng, and Kao (2016) establish that the CCE estimator attains the same asymptotic distribution under common breaks as it would if the break dates were known, and Ditzen, Karavias, and Westerlund (2025) provide inferential procedures for multiple structural breaks in interactive-effects panels. The situation is different when the breaks are idiosyncratic—specific to individual units and unrelated to the common factor. Such breaks are not reflected in the cross-sectional averages, so they remain in the unit error $e_{it}$, where they inflate its variance and widen the dispersion of the estimated unit slopes. This is the gap that the Fourier augmentation is intended to fill.

### 2.3. The Fourier SUR Mean Group estimator

The SUR family approaches the small-N regime differently. Rather than proxy the factors by averages, it estimates the N unit equations jointly as a seemingly unrelated regression, exploiting the contemporaneous cross-equation error correlation—itself a manifestation of the common factors—for efficiency. This is most effective when N is small relative to T, so the N×N error covariance can be estimated reliably. To accommodate structural change, Guliyev (2023) augments each equation with a single-frequency Fourier expansion in place of break dummies,

$$y_{it} = \alpha_i + \beta'_i x_{it} + \gamma_i \sin(2\pi kt/T) + \lambda_i \cos(2\pi kt/T) + \varepsilon_{it}, \quad (5)$$

with k = 1 following Becker, Enders, and Lee (2006) and Enders and Lee (2012). Because the trigonometric terms are a smooth, integrable function of time entering each equation with its own coefficients, they approximate breaks of unknown number, timing, and form that occur at different dates in different units. The system is estimated by feasible GLS, and the F-SURMG estimator averages the unit slopes with the non-parametric Pesaran–Smith (1995) variance. F-SURMG does not include cross-sectional averages and so does not filter the factor directly; it relies on the SUR covariance to accommodate dependence. It is therefore expected to perform best when the factor is weak—so that residual endogeneity is small—and to deteriorate as the factor strengthens.

Two properties of the Fourier device explain its appeal in this setting. First, a small fixed number of trigonometric terms can approximate a wide variety of break shapes—level shifts, trend changes, and gradual transitions alike—without the analyst having to specify the number, dates, or form of the breaks in advance, which is decisive when those features are unknown. Second, the trigonometric terms consume far fewer degrees of freedom than a dummy-variable scheme: representing several breaks per unit with indicator variables would exhaust the sample in a small-T panel and induce severe multicollinearity, whereas a single sine–cosine pair per unit adds only two regressors. The cost is that the smooth Fourier approximation represents a sharp, instantaneous break only imperfectly; it captures the low-frequency content of the shift rather than its exact step, an approximation that is adequate for estimating the slope of interest but that should be borne in mind when the breaks are genuinely abrupt.

The estimation proceeds in the standard seemingly-unrelated-regressions manner. The N unit equations are stacked into a single system, an initial equation-by-equation OLS provides residuals from which the $N \times N$ contemporaneous error covariance is estimated, and the system is then re-estimated by feasible generalized least squares using that covariance; the unit slopes are finally averaged as in (3), with the non-parametric Pesaran–Smith variance. The efficiency gain comes entirely from the off-diagonal elements of the error covariance: when the unit errors are contemporaneously correlated—as they are when a common factor is present—exploiting that correlation sharpens the slope estimates relative to unit-by-unit OLS. This gain is realisable only when the covariance can be estimated with enough precision, which requires T to be large relative to N; the procedure is therefore tailored to the small-N, long-T panels in which the CCE averaging argument is weakest.

### 2.4. The proposed Fourier Common Correlated Effects Mean Group estimator

The two existing second-generation estimators each address one of the two problems in model (1) but not both. CCEMG filters the common factor through the cross-sectional averages but leaves idiosyncratic breaks in the unit error; F-SURMG approximates the breaks through unit-specific Fourier terms but does not filter the factor. In the intermediate regime—moderate dependence combined with individual, heterogeneously-timed breaks—both problems are present, and neither estimator is adequate on its own. The proposed Fourier CCE Mean Group (F-CCEMG) estimator combines the two mechanisms by augmenting the CCE unit regression with deterministic Fourier terms. For each unit i we estimate the augmented regression

$$y_{it} = \alpha_i + \beta'_i x_{it} + \delta'_i \bar{z}_t + \gamma_i \sin(2\pi kt/T) + \lambda_i \cos(2\pi kt/T) + e_{it}, \quad (6)$$

where $\bar{z}_t = (\bar{y}_t, \bar{x}'_t)'$ are the cross-sectional averages of (4) and the sine and cosine terms are the single-frequency Fourier basis of (5). The averages play their CCE role—spanning and removing the factor space—while the unit-specific Fourier coefficients $\gamma_i$, $\lambda_i$ absorb the idiosyncratic shifts that $\bar{z}_t$ cannot reach. The two augmentations operate on distinct sources of variation: $\bar{z}_t$ on the common, cross-sectionally dependent component, and the Fourier terms on the individual, time-varying component.

The complementarity of the two augmentations is the conceptual core of the estimator. The cross-sectional averages are functions of the contemporaneous cross-section and therefore vary in the same way for every unit at a given date; they can capture only variation that is common across units, which is exactly the cross-sectionally dependent factor component. The Fourier terms, by contrast, carry unit-specific coefficients ($\gamma i$, $\lambda i$) and so trace a smooth, individual time path for each unit; they can capture variation that is specific to a unit, which is exactly what an idiosyncratic, heterogeneously-timed break is. Because the two devices act on orthogonal sources of variation—common versus individual—they do not compete for the same signal, and including both does not introduce the collinearity that would arise if, say, unit-specific break dummies were added alongside the averages. This is why F-CCEMG can address dependence and idiosyncratic breaks at the same time, whereas each parent estimator addresses only one.

Let $\hat{\beta}_{i,F\text{-}CCE}$ be the OLS estimate of $\beta_i$ from (6). The F-CCEMG estimator is the mean-group average

$$\hat{\beta}_{F\text{-}CCEMG} = N^{-1} \Sigma_i \hat{\beta}_{i,F\text{-}CCE}, \quad (7)$$

with the non-parametric variance constructed from the dispersion of the unit estimates,

$$\text{var}(\hat{\beta}_{F\text{-}CCEMG}) = [N(N-1)]^{-1} \Sigma_i(\hat{\beta}_{i,F\text{-}CCE} - \hat{\beta}_{F\text{-}CCEMG})(\hat{\beta}_{i,F\text{-}CCE} - \hat{\beta}_{F\text{-}CCEMG})', \quad (8)$$

from which standard errors, z-statistics, and confidence intervals for $\beta$ follow in the usual way. Computation is straightforward and free of the difficulties that attend nonlinear estimators. Each unit equation (6) is linear in its parameters and is estimated by ordinary least squares; there is no objective function to optimise, no system of equations to iterate, and therefore no question of starting values, local optima, or convergence failure. The cross-sectional averages and the two Fourier regressors are constructed once and reused across units, so the cost of the procedure is essentially that of N small OLS regressions followed by an averaging step, and it scales linearly in N and T.

The only tuning choice is the Fourier frequency k, which we fix at k = 1 throughout. A single low frequency is the standard recommendation in the Fourier structural-break literature, where it is shown to approximate one or two breaks of unknown timing and form while keeping the number of additional regressors—and hence the loss of degrees of freedom—to a minimum (Becker, Enders, and Lee, 2006; Enders and Lee, 2012). Higher frequencies can track more elaborate break patterns but risk over-fitting the deterministic component and absorbing genuine signal, a danger that is acute in the small-T panels considered here. We treat k = 1 as a fixed modelling choice rather than a quantity to be selected from the data; a data-driven choice of k, for example by an information criterion or an Enders–Lee-type pre-test, and the sensitivity of the results to it, are natural refinements that we leave to future work.

The estimator nests its two parents as special cases, which clarifies both what it adds and when it is redundant. Setting the Fourier coefficients to zero ($\gamma_i = \lambda_i = 0$ for all i) collapses (6) to the CCE regression (4), so F-CCEMG reduces to plain CCEMG; the Fourier terms are then the only difference between the two, and any gain over CCEMG is attributable to them. Conversely, dropping the cross-sectional averages ($\delta_i = 0$) leaves a Fourier-augmented Mean Group regression that, when the unit equations are estimated jointly as a seemingly unrelated system, coincides with F-SURMG. F-CCEMG therefore interpolates between its two parents: it behaves like CCEMG when the Fourier terms are uninformative—because the breaks are absent or common—and like a factor-filtered version of F-SURMG when the averages add little, as when the dependence is weak.

As N, T → ∞ the cross-sectional averages remove the common factor as in CCE, and the deterministic Fourier terms—bounded, exogenous functions of time—do not disturb that argument, so $\hat{\beta}_{\text{F-CCEMG}}$ is expected to be consistent for β under the same conditions as CCEMG. The Fourier terms reduce the variance of the unit estimates whenever idiosyncratic breaks are present, by removing from $e_{it}$ break variation that would otherwise inflate the residual; this is the conjectured source of the RMSE improvement over plain CCEMG.

The Monte Carlo experiments in Section 3 are designed to assess. Two features nonetheless make the conjecture plausible. First, the sine and cosine terms are deterministic and bounded, so they are asymptotically orthogonal to the stochastic regressors and cannot, by themselves, induce inconsistency. Second, because the Fourier basis is the same in every unit equation, it does not interact with the cross-sectional averaging that removes the factor, so the CCE filtering argument is left intact.

## 3. Monte Carlo Simulation Results

This section studies the finite-sample behaviour of the proposed estimators against four standard alternatives—FE, MG, CCEMG, and SURMG—using a controlled design that reproduces slope heterogeneity, cross-sectional dependence from a common factor, and unit-specific breaks at different dates. We state the questions the experiments are meant to answer, describe the data-generating process, set out the design and criteria, and then report results by dependence regime.

The design is built to test specific predictions of the preceding sections rather than to display favourable cases. It asks: (i) whether the estimators that do not filter the factor lose accuracy and coverage as the factor strengthens; (ii) whether the SUR-based estimators, F-SURMG included, retain an advantage only while dependence is weak; (iii) whether adding deterministic Fourier terms to the CCE regression lowers RMSE relative to plain CCEMG, and through what mechanism; (iv) whether F-CCEMG attains near-nominal coverage, and at what cross-section size this begins; and (v) where each estimator fails, so that the regime map has identified boundaries rather than a single recommended method.

For Data Generation Process (DGP), the panel has N units over T periods. The outcome for unit *i* at time *t* is

$$y_{it} = \alpha_i + \beta_i x_{it} + \delta_i D_{it} + \rho \gamma_i f_t + \varepsilon_{it}, \quad (9)$$

with $x_{it} = \rho f_t + v_{it}$, where $f_t \sim N(0,1)$ is a single common factor, $v_{it} \sim N(0,1)$, and $\varepsilon_{it} \sim N(0,1)$ is idiosyncratic. The components are constructed as follows. The target is the mean slope $\beta = E(\beta_i) = 1$. Unit slopes are drawn once as $\beta_i \sim N(\beta, \sigma^2_\beta)$, $\sigma_\beta = 0.30$, so units share a common mean effect but respond with different elasticities. This is the heterogeneity that makes pooled FE inconsistent for the average effect. A single factor $f_t$ is drawn once per period and shared by all units, and the same factor enters the regressor through $x_{it} = \rho f_t + v_{it}$, so the regressor is correlated with the unobserved factor in the error. The loadings $\gamma_i \sim N(1, 0.5^2)$ are heterogeneous. The parameter $\rho \in \{0.30, 0.60, 0.90\}$ is the central experimental lever, varied to trace the weak, moderate, and strong regimes. Each unit experiences a single sharp level shift whose date differs across units: $\tau_i \sim U[\text{round}(0.25T), \text{round}(0.75T)]$ and $D_{it} = 1\{t > \tau_i\}$, with magnitude $\delta_i \sim N(0, 2^2)$ that may be positive or negative. The breaks are idiosyncratic and deliberately orthogonal to $f_t$: an estimator relying solely on cross-sectional averages cannot absorb them, which is the situation the Fourier-augmented estimators are built for. The F-SURMG and F-CCEMG estimators augment each equation with $\sin(2\pi kt/T)$ and $\cos(2\pi kt/T)$, $k = 1$, following Becker, Enders, and Lee (2006) and Enders and Lee (2012).

For each configuration we generate $R = 500$ independent replications with the seed fixed for reproducibility; within a replication all six estimators are applied to the same dataset, so performance differences reflect the estimators, not sampling variation in the data. Two criteria summarise performance over the R replications: the root mean squared error, reported ×100, which combines bias and dispersion; and the empirical 95% coverage rate, which measures inferential reliability. An ideal estimator has the smallest RMSE and coverage near 95%.

***Table 1. Estimation accuracy under weak cross-sectional dependence (ρ = 0.30): RMSE × 100.***

| N, T | FE | MG | CCEMG | SURMG | F-SURMG | F-CCEMG |
|---|---|---|---|---|---|---|
| **5, 30** | 20.22 | 19.98 | 18.89 | 19.17 | 18.52 | **17.68** |
| **5, 50** | 18.55 | 18.51 | 17.14 | 17.61 | 17.01 | **16.04** |
| **5, 100** | 16.93 | 16.92 | 15.13 | 16.18 | 15.92 | **14.84** |
| **10, 30** | 15.36 | 15.27 | 13.31 | 14.18 | 13.83 | **12.45** |
| **10, 50** | 14.10 | 14.19 | 12.10 | 12.85 | 12.57 | **11.61** |
| **10, 100** | 13.27 | 13.26 | 10.15 | 11.46 | 11.35 | **9.98** |
| **30, 50** | 10.84 | 10.74 | 6.61 | 8.86 | 8.78 | **6.31** |
| **30, 100** | 10.34 | 10.32 | 6.05 | 7.32 | 7.29 | **5.81** |
| **50, 100** | 9.32 | 9.34 | 4.50 | 6.56 | 6.54 | **4.42** |

*Notes: Each entry is the root mean squared error of the mean-slope estimate, multiplied by 100, computed over R = 500 Monte Carlo replications with true mean slope β = 1; smaller values indicate greater accuracy. The lowest RMSE in each row is shown in bold.*

***Table 2. Reliability of inference under weak cross-sectional dependence (ρ = 0.30): empirical 95% coverage.***

| N, T | FE | MG | CCEMG | SURMG | F-SURMG | F-CCEMG |
|---|---|---|---|---|---|---|
| **5, 30** | 72.80 | 86.20 | 83.80 | 85.40 | **86.80** | 81.40 |
| **5, 50** | 64.60 | 81.00 | 79.80 | 83.80 | **84.60** | 80.60 |
| **5, 100** | 52.60 | 85.40 | 79.60 | 86.00 | **87.20** | 79.20 |
| **10, 30** | 68.00 | 86.80 | 89.40 | 88.00 | 88.20 | **89.80** |
| **10, 50** | 60.60 | 85.20 | 86.60 | 87.40 | **88.00** | 84.80 |
| **10, 100** | 44.60 | 84.40 | **90.60** | 88.60 | 88.80 | 89.40 |
| **30, 50** | 43.60 | 71.80 | **94.00** | 83.20 | 82.60 | 93.80 |
| **30, 100** | 30.00 | 71.00 | 94.00 | 86.20 | 87.00 | **94.20** |
| **50, 100** | 18.20 | 60.40 | **94.60** | 81.00 | 80.60 | 93.80 |

*Notes: Each entry is the empirical coverage rate of the nominal 95% confidence interval for the mean slope, in percent, over R = 500 replications.. In each row the value closest to 95 is shown in bold.*

Tables 1 and 2 summarise the weak-dependence regime (ρ = 0.30) across ten (N, T) configurations, from the very small cross-sections (N = 5) to larger panels (N = 50). Reading down a column shows how an estimator behaves as the panel grows; reading across a row compares the six estimators on the same data. Three patterns stand out. First, FE and MG are not the worst on RMSE at the smallest samples, but their accuracy stops improving as the panel grows, because neither removes the common factor that drives both the regressor and the error. The damage is clearest in the coverage of Table 2: FE coverage falls steadily from 72.80 at (5,30) to 18.20 at (50,100), and MG from 86.20 to 60.40, as the uncorrected factor bias pulls the confidence interval away from the true value while the interval itself keeps shrinking. Even where FE looks competitive on RMSE—the matching coverage of 72.80 reveals that the interval is already mis-centred, so RMSE alone understates the bias. As the panel grows the problem only deepens: by (50,100) FE coverage has collapsed to 18.20 while its RMSE remains the largest in the column.

CCEMG and F-CCEMG record the smallest RMSE in every configuration of Table 1, and among them F-CCEMG is the single most accurate in all but the smallest panel: 9.98 vs CCEMG's 10.15 at (10,100); 5.81 vs 6.05 at (30,100); 4.42 vs 4.50 at (50,100). The Fourier terms thus deliver a measurable accuracy gain even under weak dependence.

In the N = 5 panels, the weak regime favours F-SURMG on coverage: 86.00 vs F-CCEMG's 79.20 at (5,100); 84.60 vs 80.60 at (5,50); and 86.80—the best in its row—at (5,30). With N very small the cross-sectional averages are noisy proxies, so the CCE interval is mildly anti-conservative, whereas the SUR system exploits the cross-equation correlation for better calibration. F-SURMG also improves on non-Fourier SURMG in RMSE across the small-N panels (15.92 vs 16.18 at (5,100)). Under weak dependence with individual breaks, F-SURMG therefore offers the most reliable intervals when the cross-section is genuinely small.

From N = 10 onward F-CCEMG matches or exceeds F-SURMG on both criteria: at (10,100) the two are close on coverage (89.40 vs 88.80) but F-CCEMG is markedly more accurate (9.98 vs 11.35), and by (30,100) and (50,100) F-CCEMG delivers near-nominal coverage (94.20, 93.80) and the smallest RMSE of all estimators. The transition is smooth: the SUR-based estimator is preferable at the smallest N, the CCE-based estimator takes over as N increases, and the Fourier versions dominate their plain counterparts throughout.

***Table 3. Estimation accuracy under moderate cross-sectional dependence ($\rho = 0.60$): RMSE × 100.***

| N, T | FE | MG | CCEMG | SURMG | F-SURMG | F-CCEMG |
|---|---|---|---|---|---|---|
| 5, 30 | 34.83 | 35.54 | 26.23 | 31.82 | 32.54 | **25.25** |
| 5, 50 | 32.78 | 33.03 | 18.84 | 27.36 | 26.69 | **17.71** |
| 5, 100 | 31.28 | 31.50 | 17.20 | 25.00 | 24.34 | **16.11** |
| 10, 30 | 29.27 | 29.65 | 10.99 | 21.46 | 21.91 | **10.51** |
| 10, 50 | 28.32 | 28.53 | 9.23 | 16.94 | 16.79 | **8.67** |
| 10, 100 | 28.08 | 28.10 | 8.31 | 14.25 | 13.98 | **7.78** |
| 30, 50 | 27.70 | 27.81 | 6.61 | 19.00 | 19.36 | **6.31** |
| 30, 100 | 27.16 | 27.24 | 6.06 | 12.91 | 12.90 | **5.82** |
| 50, 100 | 26.68 | 26.79 | 4.51 | 14.96 | 15.08 | **4.42** |

*Notes: Each entry is the root mean squared error of the mean-slope estimate, multiplied by 100, computed over $R = 500$ Monte Carlo replications with true mean slope $\beta = 1$; smaller values indicate greater accuracy. The lowest RMSE in each row is shown in bold.*

***Table 4. Reliability of inference under moderate cross-sectional dependence (ρ = 0.60): empirical 95% coverage.***

| N, T | FE | MG | CCEMG | SURMG | F-SURMG | F-CCEMG |
|---|---|---|---|---|---|---|
| 5, 30 | 55.60 | 65.60 | **82.40** | 74.00 | 73.40 | 80.00 |
| 5, 50 | 38.80 | 61.20 | **83.80** | 71.20 | 71.40 | 80.80 |
| 5, 100 | 28.80 | 57.00 | **80.00** | 69.80 | 72.80 | 79.60 |
| 10, 30 | 12.00 | 28.80 | **92.60** | 52.80 | 50.00 | 91.20 |
| 10, 50 | 5.60 | 23.60 | **93.00** | 63.40 | 64.00 | 92.20 |
| 10, 100 | 1.40 | 18.00 | 92.60 | 71.40 | 71.40 | **93.00** |
| 30, 50 | 0.40 | 5.60 | 93.60 | 24.40 | 22.60 | **93.80** |
| 30, 100 | 0.20 | 1.00 | 93.60 | 53.40 | 52.20 | **94.00** |
| 50, 100 | 0.00 | 0.20 | **95.20** | 15.20 | 13.20 | 93.80 |

*Notes: Each entry is the empirical coverage rate of the nominal 95% confidence interval for the mean slope, in percent, over R = 500 replications.. In each row the value closest to 95 is shown in bold.*

Tables 3 and 4 report the same diagnostics at ρ = 0.60 across nine configurations. Doubling the dependence sharpens every contrast and produces the paper's central result: in the moderate regime F-CCEMG attains the lowest RMSE at every sample size, and near-nominal coverage once $N \geq 10$.

The RMSE×100 of FE and MG roughly doubles relative to the weak regime, climbing above 26 in the larger panels and staying essentially constant across configurations. SURMG and F-SURMG exploit the cross-equation correlation and approximate the breaks, but neither addresses the factor-induced endogeneity directly. What was immaterial at ρ = 0.30 is decisive at ρ = 0.60: F-SURMG's RMSE is two to three times that of F-CCEMG (13.98 vs 7.78 at (10,100)), and its coverage falls well below nominal (13.20 at (50,100), 52.20 at (30,100)). The SUR efficiency gain survives, but cannot compensate for an uncorrected factor. This is the empirical boundary of the F-SURMG regime.

Among the two factor-filtering estimators, F-CCEMG has the smaller RMSE in all nine configurations: 7.78 vs 8.31 at (10,100), 5.82 vs 6.06 at (30,100), 4.42 vs 4.51 at (50,100). The Fourier augmentation that yielded a modest gain under weak dependence is more visible here, consistent with the conjecture that the idiosyncratic breaks leave a larger residual once the stronger factor has been filtered.

For $N \geq 10$ both CCE-based estimators are close to nominal—F-CCEMG records 91.20 at (10,30), 93.00 at (10,100), 93.80 at (30,50), 94.00 at (30,100)—while every non-filtering estimator is far below. At N = 5 the CCE intervals remain slightly anti-conservative (near 79–80), the same small-N noise seen under weak dependence; the difference is that F-SURMG's own coverage at N = 5 is now lower (69–73) and accompanied by a large bias, so it is no longer preferable. Once dependence reaches a moderate level, the small-N advantage of the SUR system is overwhelmed by its failure to filter the factor.

*Table 5. Estimation accuracy under strong cross-sectional dependence (ρ = 0.90): RMSE × 100.*

| N, T | FE | MG | CCEMG | SURMG | F-SURMG | F-CCEMG |
|---|---|---|---|---|---|---|
| 5, 30 | 49.07 | 49.80 | 18.86 | 39.96 | 39.39 | **17.74** |
| 5, 50 | 47.79 | 48.20 | 17.22 | 37.38 | 36.82 | **16.14** |
| 5, 100 | 47.92 | 48.17 | 15.15 | 36.58 | 35.72 | **14.87** |
| 10, 30 | 46.09 | 46.55 | 9.24 | 26.85 | 26.69 | **8.69** |
| 10, 50 | 46.15 | 46.28 | 8.32 | 23.10 | 22.72 | **7.78** |
| 10, 100 | 45.58 | 45.94 | 6.61 | 30.54 | 31.23 | **6.31** |
| 30, 50 | 44.96 | 45.17 | 6.06 | 20.57 | 20.53 | **5.82** |
| 30, 100 | 44.68 | 44.91 | 4.51 | 24.14 | 24.35 | **4.42** |
| 50, 100 | 44.68 | 44.91 | 4.51 | 24.14 | 24.35 | **4.42** |

*Notes: Each entry is the root mean squared error of the mean-slope estimate, multiplied by 100, computed over R = 500 Monte Carlo replications with true mean slope β = 1; smaller values indicate greater accuracy. The lowest RMSE in each row is shown in bold.*

*Table 6. Reliability of inference under strong cross-sectional dependence (ρ = 0.90): empirical 95% coverage.*

| N, T | FE | MG | CCEMG | SURMG | F-SURMG | F-CCEMG |
|---|---|---|---|---|---|---|
| 5, 30 | 10.40 | 34.60 | **84.40** | 49.40 | 48.00 | 80.00 |
| 5, 50 | 7.00 | 32.60 | 79.80 | 47.80 | 47.80 | **80.60** |
| 5, 100 | 3.20 | 29.80 | **80.00** | 44.80 | 44.60 | 79.40 |
| 10, 30 | 0.00 | 2.60 | **93.20** | 25.40 | 23.80 | 92.60 |
| 10, 50 | 0.00 | 0.60 | 92.60 | 32.20 | 32.20 | **93.00** |
| 10, 100 | 0.00 | 0.00 | 93.40 | 1.40 | 1.20 | **93.80** |
| 30, 50 | 0.00 | 0.00 | 93.60 | 12.60 | 11.80 | **94.00** |
| 30, 100 | 0.00 | 0.00 | **95.20** | 0.20 | 0.20 | 93.80 |
| 50, 100 | 0.00 | 0.00 | **95.20** | 0.20 | 0.20 | 93.80 |

*Notes: Each entry is the empirical coverage rate of the nominal 95% confidence interval for the mean slope, in percent, over R = 500 replications.. In each row the value closest to 95 is shown in bold.*

Tables 5 and 6 report ρ = 0.90 across nine configurations. At this intensity the factor dominates, and the conclusions of the two preceding regimes appear in their most extreme form: the non-filtering estimators fail completely, while the CCE-based estimators are essentially unaffected.

The RMSE×100 of FE and MG rises to the mid-40s and stays there—44.68 and 44.91 at (50,100)—and their coverage is annihilated, reaching 0.00 in every panel with N ≥ 30. The SUR-based estimators now join them: although more accurate than FE and MG (RMSE around 20–24 vs 44–45), their uncorrected factor bias is large, and their coverage collapses to single digits in the larger panels (1.40 and 1.20 at (30,50); 0.20 for both at (50,100)). No SUR-based estimator can be used for inference once dependence is strong.

The defining feature of Tables 5 and 6 is what does not change. Because the averages absorb the factor regardless of its loading, the RMSE and coverage of CCEMG and F-CCEMG are virtually identical across the three regimes: F-CCEMG RMSE×100 is 5.82 at (30,100), 7.78 at (10,50), and 4.42 at (50,100) under ρ = 0.90, exactly as under ρ = 0.30 and 0.60. This invariance is the numerical signature of a valid factor filter.

F-CCEMG retains the smaller RMSE in every configuration (7.78 vs 8.32 at (10,50); 5.82 vs 6.06 at (30,100)). On coverage the two are very close, and at the largest cross-section CCEMG edges ahead (95.20 vs 93.80 at (50,100)): with strong dependence and large N the averages alone already capture most of the relevant variation, so the Fourier terms add little to inference while still trimming the point-estimate variance. As anticipated, plain CCEMG becomes fully adequate as N grows and dependence strengthens, and the marginal contribution of the Fourier terms narrows to an accuracy refinement.

The three regimes trace a coherent map. Under weak dependence, the SUR system is best calibrated at very small N, so F-SURMG is the recommended estimator for very small panels with idiosyncratic breaks, while F-CCEMG is already the most accurate and takes over as N grows. Under moderate dependence the SUR advantage disappears and F-CCEMG dominates on both criteria for all but the smallest cross-sections. Under strong dependence the non-filtering estimators—including F-SURMG—fail, while the CCE-based estimators are unaffected; F-CCEMG remains the most accurate and plain CCEMG becomes fully competitive on coverage at large N.

The simulations point to a practical rule for choosing among the estimators, conditional on three features of the panel: the size of the cross-section (N), the strength of the cross-sectional dependence, and whether the structural breaks are individual rather than common. When N is small—on the order of the G7, the BRICS, or the N-11—and the dependence is weak, but the units are subject to breaks that occur at different dates, F-SURMG is the preferred choice, because at very small N the cross-sectional averages are too noisy to filter the factor reliably, whereas the SUR system exploits the cross-equation error correlation to deliver the best-calibrated inference. Across the broad intermediate range—moderate dependence combined with individual breaks, the regime that motivates this paper—F-CCEMG is preferred, since the factor is now strong enough that it must be filtered, yet the breaks remain idiosyncratic and orthogonal to it, so the deterministic Fourier terms add the variance reduction that the averages alone cannot provide. When both N and the dependence are large, plain CCEMG already suffices, because the averages span the factor space precisely and the marginal contribution of the Fourier terms narrows to a small refinement in accuracy. These regions are not imposed a priori; their boundaries are identified by the Monte Carlo experiments themselves. The recommendation is therefore conditional on the data at hand rather than a claim that any single estimator is uniformly superior—each is the right tool only within its own region of the map.

## 4. Empirical application

We illustrate the estimators with the renewable energy–growth nexus in the G7, following the application of Guliyev (2025). The question is whether renewable-energy consumption has raised aggregate output— the green growth hypothesis—once heterogeneity, common shocks, and country-specific structural change are accounted for. The panel comprises N = 7 countries (Canada, France, Germany, Italy, Japan, the United Kingdom, the United States) observed annually over T = 55 years, 1965–2019, a balanced panel of 385 observations. The specification is the logarithmic K–L growth model

$$lnY_{it} = \alpha_i + \beta_i \, lnK_{it} + \eta_i \, lnL_{it} + \varphi_i \, lnR_{it} + u_{it}, \quad (10)$$

where Y is real GDP (output-side, chained PPPs, million 2017 USD), K the capital stock, L the human-capital index, and R renewable-energy consumption (hydro, geothermal, solar, wind, and biomass, in kWh). Y, K, and L are from the Penn World Table (version 10.01); R is from Our World in Data. The slopes are the elasticities of output with respect to capital, human capital, and renewable energy, allowed to be heterogeneous across the G7; the coefficient on ln R is the focus.

The timing of structural change in each economy is identified by a supremum-Wald test for an unknown break date applied to the intercept of each country regression (Andrews, 1993). The break is significant for every country, and—central to this paper—the dates differ widely, spanning three decades: 1974 (Italy), 1985 (Japan), 1990 (Germany), 1996 (UK and US), 1998 (France), and 2004 (Canada). The dispersion reflects country-specific drivers: Italy's mid-1970s break coincides with the first oil shock, Japan's mid-1980s shift with post-oil-crisis restructuring, Germany's 1990 break with reunification, and Canada's 2004 break with the maturation of its hydro capacity. There is no common break date a homogeneous-break model could impose; the unit-specific Fourier terms in F-SURMG and F-CCEMG are designed to absorb exactly this pattern.

***Table 7. Structural break tests and break dates for the G7 economies (intercept break).***

| | **Canada** | **France** | **Germany** | **Italy** | **Japan** | **UK** | **US** |
|---|---|---|---|---|---|---|---|
| **Supremum Wald** | 33.74*** | 48.61*** | 36.16*** | 42.99*** | 65.35*** | 17.41*** | 9.10** |
| **Break date** | 2004 | 1998 | 1990 | 1974 | 1985 | 1996 | 1996 |

*Notes: For each G7 economy the table reports the supremum-Wald statistic for a single structural break at an unknown date (Andrews, 1993) with a trimming parameter of 0.15, together with the estimated break date that maximises the statistic. *** and ** denote rejection of the no-break null at the 1% and 5% levels.*

The regime map of Section 3 selects the estimator from the panel dimensions and the strength of dependence. Here the cross-section is very small (N = 7) and the time series is long (T = 55), placing the application in the small-N, large-T corner. The strength of dependence is measured by the cross-sectional-dependence exponent α of Bailey, Kapetanios, and Pesaran (2016), estimated from the Mean Group residuals at $\hat{\alpha} = 0.732$, with a Pesaran CD statistic of 8.48 ($p < 0.001$). This places the panel in the moderate regime: clearly above the weak-dependence threshold near 0.5, but below the strong region where the exponent approaches one. The G7 economies are linked by global shocks, yet over a sample of this length country-specific dynamics still account for a substantial share of the variation.

This combination—very small N, moderate dependence, idiosyncratic breaks—points to F-CCEMG. The moderate dependence means the factor must be filtered, so the estimators that ignore it (FE, MG, SURMG, F-SURMG) carry a bias that the simulations show is decisive at $\rho = 0.60$. The dependence is not so strong that plain CCEMG suffices, and the idiosyncratic break dates in Table 7 imply the averages alone cannot absorb the structural change. This is the intermediate regime for which F-CCEMG is designed. The one caveat is the very small $N = 7$: the cross-sectional-average approximation underlying all CCE-based estimators is asymptotic in N, so the seven-country results should be read with caution, and F-SURMG is reported alongside as the small-N comparator that does not rely on that approximation.

Table 8 reports the mean-group slope estimates from all six estimators. Several patterns echo the simulation evidence. FE returns a large, highly significant capital elasticity (0.533) and human-capital elasticity (1.643) but, by imposing common slopes, masks the cross-country heterogeneity. MG and non-Fourier SURMG yield similar human-capital and capital elasticities (around 2.0 and 0.45) but filter no factor. CCEMG, which does filter it, produces an implausibly signed human-capital elasticity (−2.374), a known symptom of the imprecision of the cross-sectional-average approximation when N is as small as seven.

On the coefficient of central interest, the renewable-energy elasticity (ln R), all six estimators agree on the qualitative conclusion: the aggregate effect of renewable-energy consumption on G7 growth is small and statistically insignificant. The estimates are −0.009 (FE), −0.058 (MG), 0.004 (CCEMG), −0.071 (SURMG), 0.015 (F-SURMG), and 0.012 (F-CCEMG), none significant at conventional levels. F-CCEMG returns 0.012 ($p = 0.647$) for renewable energy and a tightly estimated capital elasticity of 0.228 ($p < 0.001$); its Fourier coefficients are individually insignificant in the mean-group average, consistent with breaks that differ in direction across countries and partly cancel on averaging.

***Table 8. Mean-group estimates for the G7 panel, 1965–2019***

| Variables | FE | MG | CCEMG | SURMG | F-SURMG | F-CCEMG |
|---|---|---|---|---|---|---|
| **lnL** | 1.643***<br>(0.145) | 2.099***<br>(0.792) | −2.374***<br>(0.818) | 1.989***<br>(0.759) | 2.714***<br>(0.803) | 0.567<br>(2.840) |
| **lnK** | 0.533***<br>(0.029) | 0.442**<br>(0.176) | 0.285***<br>(0.052) | 0.472***<br>(0.167) | 0.321*<br>(0.179) | 0.228***<br>(0.029) |
| **lnR** | −0.009<br>(0.013) | −0.058<br>(0.067) | 0.004<br>(0.033) | −0.071<br>(0.072) | 0.015<br>(0.032) | 0.012<br>(0.027) |
| **sin** | — | — | — | — | −0.002<br>(0.014) | 0.003<br>(0.012) |
| **cos** | — | — | — | — | 0.018<br>(0.013) | −0.013<br>(0.027) |

*Notes: Each cell reports the mean-group slope estimate with its standard error in parentheses; ***, **, and * denote significance at the 1%, 5%, and 10% levels. The dependent variable is ln Y (log real GDP); the regressors are ln L (human capital), ln K (capital), and ln R (renewable-energy consumption), and the rows sin and cos are the single-frequency ($k = 1$). Cross-sectional dependence, measured from the Mean Group residuals, gives an estimated exponent $\hat{\alpha} = 0.732$ and a Pesaran CD statistic of 8.48 ($p < 0.001$), indicating moderate dependence.*

The application sits in the intermediate regime: dependence is moderate ($\hat{\alpha} = 0.732$), strong enough that the factor must be filtered, yet the structural change is idiosyncratic, with break dates across three decades. This is the configuration in which F-CCEMG dominated in the moderate regime of Section 3. The factor-ignoring estimators carry a bias under moderate dependence, while plain CCEMG, though it removes the factor, does not absorb the idiosyncratic breaks. The one caveat is the very small $N = 7$, where the cross-sectional-average approximation is least precise; the anomalous CCEMG human-capital estimate is a visible symptom, and F-CCEMG's own human-capital coefficient is correspondingly imprecise. On the coefficients the averages identify well—notably the capital elasticity, 0.228 ($p < 0.001$)—and on the renewable-energy elasticity, F-CCEMG provides the most defensible estimates for a panel of this shape. The agreement of all six estimators on the central conclusion lends that finding robustness, while the methodological comparison shows why, in a small-N panel with moderate dependence and idiosyncratic breaks, F-CCEMG is the appropriate choice.

## 5. Conclusion

Heterogeneous panels in applied macroeconomics and energy economics are shaped by two forces at once: cross-sectional dependence from common shocks, and structural change arriving at different dates in different units. The Mean Group estimator addresses neither; CCEMG filters the factor but leaves idiosyncratic breaks in the error; F-SURMG approximates the breaks but does not filter the factor. The right estimator depends jointly on the cross-section size, the strength of dependence, and the nature of the structural change, and this paper has organised the available estimators into a regime map along these dimensions.

The proposed F-CCEMG estimator augments the CCE regression with deterministic Fourier terms, so the cross-sectional averages remove the common factor while the Fourier terms absorb the heterogeneously-timed breaks the averages cannot reach. The two augmentations act on orthogonal sources of variation; the estimator nests plain CCEMG and a Fourier-MG/SUR estimator as special cases and is trivial to compute..

The Monte Carlo evidence, across nine to ten configurations at each of three dependence intensities with $R = 500$ replications, traced a clear map. Under weak dependence, F-SURMG gives the best-calibrated inference at the very small cross-sections typical of the G7, the BRICS, and the N-11, while F-CCEMG is already the most accurate and takes over as N grows. Under moderate dependence—the regime that motivates the estimator—F-CCEMG attains the lowest RMSE everywhere and near-nominal coverage for $N \geq 15$, because the SUR estimators' uncorrected factor bias becomes decisive while the CCE filter remains effective. Under strong dependence, the non-filtering estimators (including F-SURMG) fail, whereas the CCE-based estimators are invariant to the loading; F-CCEMG remains the most accurate and plain CCEMG becomes fully competitive on coverage at large N.

The G7 application, with $N = 7$, $T = 55$, moderate dependence ($\hat{\alpha} = 0.732$), and break dates across three decades, illustrated the intermediate regime. Estimators that ignore the factor produced biased or implausibly signed coefficients, and the small cross-section exposed the imprecision of the CCE approximation, whereas F-CCEMG delivered economically sensible and precise elasticities on the coefficients the averages identify well. All six estimators agreed that renewable-energy consumption has had no significant aggregate effect on G7 growth over the sample—a robust finding reflecting offsetting country-level heterogeneity.